\documentclass[aps,prl,twocolumn,amsmath,amssymb,floatfix,nobalancelastpage]{revtex4}
\usepackage{microtype}
\usepackage{newtxtext}
\usepackage[upint,varbb]{newtxmath}
\usepackage{eucal}
\usepackage{graphicx}
\usepackage[caption=false]{subfig}
\usepackage{amsfonts}
\usepackage{amsmath}
\usepackage{amssymb}
\usepackage{textcomp}
\usepackage{color}
\usepackage{bm}
\usepackage{float}
\usepackage[colorlinks,allcolors=blue]{hyperref}
\usepackage[capitalize]{cleveref}
\usepackage{soul}

\definecolor{DarkRed}{rgb}{0.65,0,0}%
\definecolor{Green}{rgb}{0,0.5,0.3}
\definecolor{Purple}{rgb}{0.3,0,0.65}
\definecolor{DarkGray}{rgb}{0.7,0.7,0.7}
\definecolor{Blue}{rgb}{0,0,1}

\renewcommand{\vec}[1]{\bm{#1}}

\begin{document}
\title{Vortex spin-valve on a topological insulator}
\author{Morten Amundsen}
\author{Henning G. Hugdal}
\author{Asle Sudb{\o}}
\author{Jacob Linder}
\affiliation{Center for Quantum Spintronics, Department of Physics, Norwegian \\ University of Science and Technology, NO-7491 Trondheim, Norway}

\begin{abstract}
	\noindent
	Spin-valve structures are usually associated with the ability to modify the resistance of electrical currents. We here demonstrate a profoundly different effect of a spin-valve. In combination with a topological insulator and superconducting materials, we show that a spin-valve can be used to toggle quantum vortices in and out of existence. In the antiparallel configuration, the spin-valve causes superconducting vortex nucleation. In the parallel configuration, however, no vortices appear. This switching effect suggests a new way to control quantum vortices.
	\end{abstract}

\maketitle

\noindent
\emph{Introduction}. 
Topological insulators are fascinating materials which are insulating in their bulk, but have topologically protected conducting surface states~\cite{hasan_rmp_2010}. When a conventional $s$-wave superconductor is placed in contact with a topological insulator, the superconducting correlations induced on the TI surface gain a topological character~\cite{qi_rmp_2011}. This may give rise to a range of exotic phenomena, such as the appearance of Majorana bound states at vortices~\cite{fu_prl_2008}, which provides an exciting avenue towards non-Abelian statistics and topological quantum computation~\cite{nayak_rmp_2008}.  

A particularly interesting property of the surface states of a TI is the presence of spin--momentum locking. By proximity coupling both superconducting and ferromagnetic elements to the topological insulator, this may be used to create complex supercurrent density distributions \cite{zyuzin_prb_2016}. A key observation is that the exchange field enters the Hamiltonian for the surface states of a TI in the same way as the magnetic vector potential does, due to the spin-momentum locking. Because of this, one might expect that quantum vortices with a phase-winding could be induced by an exchange field alone on the surface of a TI in contact with a superconductor, without the need of any external magnetic flux. The study of superconducting vortices induced in non-superconducting materials via proximity has recently attracted attention both theoretically \cite{cuevas_prl_2007, bergeret_jltp_2008, zyuzin_prb_2016, ostroukh_prb_2016, amundsen_prl_2018} and experimentally \cite{roditchev_nphys_2015}.

In this Letter, we show that a spin-valve structure combined with a topological insulator and superconducting materials can be used to toggle quantum vortices in and out of existence. The spin-valve consists of two ferromagnetic layers which can be either in a parallel (P) or antiparallel (AP) configuration. In the P configuration, the spin-valve does not cause superconducting vortex nucleation. In contrast, vortices can exist in the AP configuration. This switching effect suggests a new way to control quantum vortices in heterostructures. The precise conditions under which this can occur will be detailed below.

\begin{figure}
\includegraphics[width=\columnwidth]{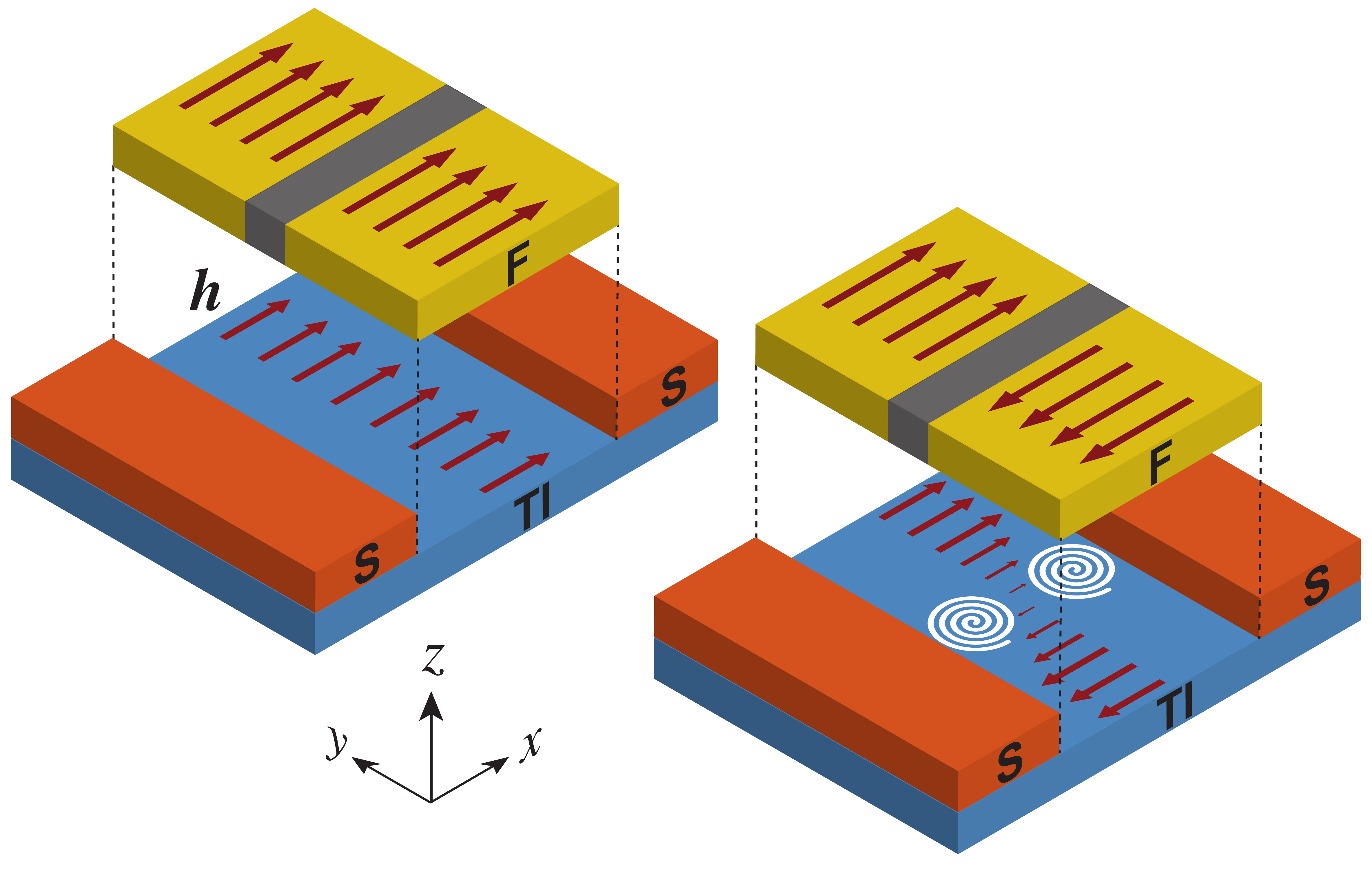}
\caption{The geometry considered. Two superconductors and a spin valve are placed on top of a topological insulator. The spatial variation of the exchange field induced by the antiparallel configuration of the ferromagnets creates vortices in the TI. The spin valve is shown lifted for clarity.}
\label{fig:geometry}
\end{figure}

To demonstrate this effect, we consider the system shown in \cref{fig:geometry}. Two superconductors are placed on top of a topological insulator, and between them is placed a pair of ferromagnets. This creates an effective SFS Josephson weak link on the two dimensional surface of the TI via the proximity effect. The distance between the superconductors is $L = 2\xi$, where $\xi$ is the superconducting coherence length, which is assumed to also be the width of the system. The exchange field in the ferromagnet is directed along the $x$ axis (between the superconductors). The magnitude of the exchange field is constant in the $x$ direction, but can be toggled between either a parallel or antiparallel configuration. Such a system can be experimentally designed by separating the two ferromagnets by a thin nonmagnetic spacer layer. If the ferromagnets have different coercive fields, one may toggle between configurations for instance by heating the system to above the critical temperature of the superconductors, $T_c$, apply a magnetic field in the $x$ direction large enough to switch the magnetization in one of the layers, and then cool the system to below $T_c$. To ensure different coercive fields, the ferromagnets may either be different materials, or have different sizes.

The surface of the three dimensional diffusive topological insulator here considered may be described by using quasiclassical theory~\cite{rammer_rmp_1986, belzig_sm_1999}. In equilibrium, all physical observables may be computed from the $2\times 2$ retarded Green function
\begin{align} 
G = \begin{pmatrix}
g & f \\
\tilde{f} & -g
\end{pmatrix},
\label{eq:GR}
\end{align}
where $g$ and $f$ are the normal and anomalous Green functions, respectively, and $\tilde{f}(\varepsilon) = f^*(-\varepsilon)$. $G$ has structure only in particle--hole space, and the spin structure has been factored out by a unitary transformation in order to take the spin--momentum locking into account. A detailed description of this procedure is given in Ref.~\cite{hugdal_prb_2017}. In the diffusive limit, the Green function is governed by the Usadel equation~\cite{usadel_prl_1970}
\begin{align}
2Di\hat{\nabla}\cdot( G\hat{\nabla} G) = \left[\varepsilon\sigma_3 , G\right],
\label{eq:usadel}
\end{align} 
where $\hat{\nabla}G = \nabla G - \frac{i}{v_F}\left[\vec{h}\sigma_3,G\right]$, $\vec{h}$ is the in-plane exchange field, $D$ is the diffusion constant, $\varepsilon$ is the quasiparticle energy, $v_F$ is the Fermi velocity and $\sigma_3$ is the third Pauli matrix. We solve \cref{eq:usadel} in the region of the TI located between the superconductors, which we consider as large enough to be described by their bulk expressions, $G_{\text{BCS}}$, as given in Ref.~\cite{hugdal_prb_2017}. We neglect the inverse proximity effect
which is a good approximation as long as the Fermi level $\mu_\mathrm{TI}$ in the TI is substantially different from $\pm\sqrt{2m v_F^2 \mu_\mathrm{S}}$, where $m$ and $\mu_\mathrm{S}$ are the electron mass and Fermi level in the superconductor, respectively \cite{hugdal_2018}. We further assume transparent boundary conditions to the superconductors, while the vacuum interfaces are described by the Neumann boundary condition $\nabla G = 0$. We note in particular that the in-plane exchange field enters \cref{eq:usadel} in precisely the same way as does the vector potential in a normal metal. A consequence of this is that the system will react to a spatial variation in $\vec{h}$ in the same way as if an effective flux  $\Phi_h = \int_A\nabla\times\vec{h}\;d\vec{r}$ is applied, where $A$ is the area of the TI surface. This means that for a sufficiently large inhomogeneous exchange field vortices may appear. Note that for a curl-free  inhomogenous $\vec{h}$, vortices do not appear. An analogy to an SNS junction with a uniform applied magnetic flux is found by considering an exchange field $\vec{h} = -h_0y\hat{x}$. In the Fraunhofer limit, where the width of the junction (in the $y$ direction) is much larger than its length, the number of vortices in the system is equal to the number of flux quanta that is applied. The relevant flux quantum for the exchange field induced vortices in the present manuscript is then $\Phi_0 = \frac{h v_F}{2}$. The square geometry of the system studied herein influences the number and position of the vortices. However, the number of flux quanta produced by the effective flux $\Phi_h$ still remains a good estimate for the number of vortices. 

From the retarded Green function, $G$, the density of states, normalized by its value at the Fermi level, may be computed as $N(\vec{r},\varepsilon) = \text{Re}\,g(\vec{r},\varepsilon)$, with $g(\vec{r},\varepsilon)$ defined in \cref{eq:GR}. Furthermore, the pair correlation in the TI, which is a measure of the strength of the superconducting correlations induced by the proximity effect, may be computed from
\begin{align}
\Psi(\vec{r}) = N_0 \int d\varepsilon\;\left[f(\vec{r},\varepsilon) - f(\vec{r},-\varepsilon)\right]\tanh\frac{\beta\varepsilon}{2},
\label{eq:PC}
\end{align}
where $\beta = 1/k_BT$, $T$ is the temperature and $N_0$ is the density of states at the Fermi level. Finally, the current density is given as
\begin{align}
\vec{J}(\vec{r}) = J_0\int d\varepsilon\;\text{Re}\left[f\nabla\tilde{f} - \tilde{f}\nabla f - \frac{4i}{v_F}\vec{h}f\tilde{f}\right]\tanh\frac{\beta\varepsilon}{2},
\label{eq:current}
\end{align}
with $J_0 = N_0eD$.

We consider an in-plane exchange field and set $\vec{h} = h_x(y) \hat x$. The necessary (but not sufficient) requirement for inducing vortices is then that $\partial_y h_x \neq 0$. To be specific, we assume that the antiparallel configuration of the ferromagnets induces an antisymmetric exchange field with a spatial variation given by $\vec{h} = h_0\tanh \left(\alpha y / L\right)\hat{x}$, where $\alpha$ is a shape factor which determines the size of the transition region. We note that the size of the effective flux $\Phi_h$, and thus the net number of vortices introduced, does not depend on the specific shape of the exchange field, since by the fundamental theorem of calculus, $\Phi_h = L\left[h(L/2) - h(-L/2)\right]$. To model the parallel configuration, a constant exchange field $\vec{h} = h_0\hat{x}$ is assumed.  

\begin{figure}
\centering
\includegraphics[width=\columnwidth]{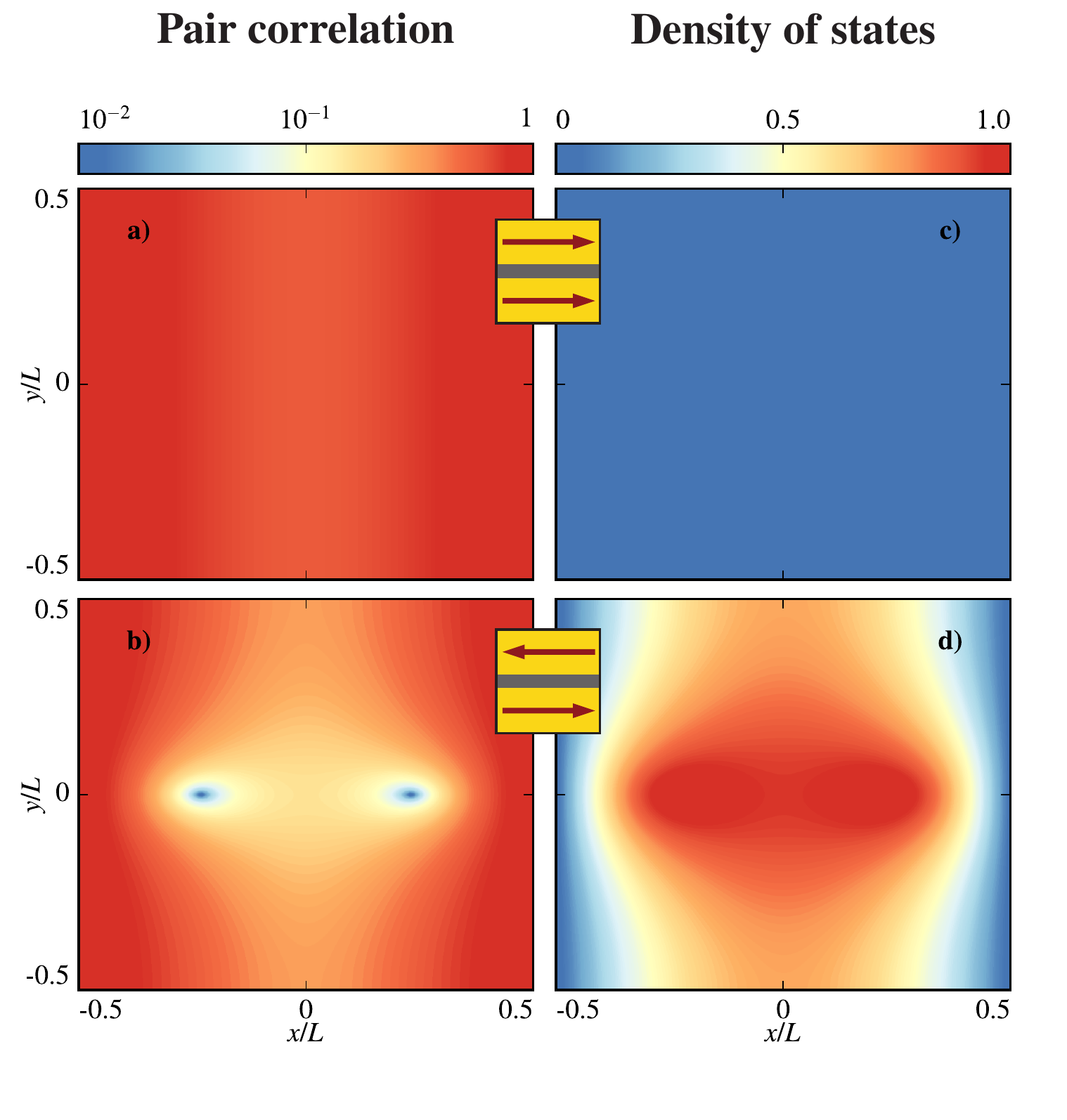}
\caption{A comparison of the results for the parallel (upper row) and antiparallel (lower row) configuration of the spin valve. a) and b) shows the absolute value of the pair correlation for the two configurations, as given by \cref{eq:PC}. The localized zeroes in the antiparallel case indicate vortices. c) and d) show the density of states $N$ at zero energy, which is gapped in the parallel configuration, and admits a normal state solution at the location of the vortices in the antiparallel configuration.}
\label{fig:vortex}
\end{figure}

\begin{figure}
\centering
\includegraphics[width=\columnwidth]{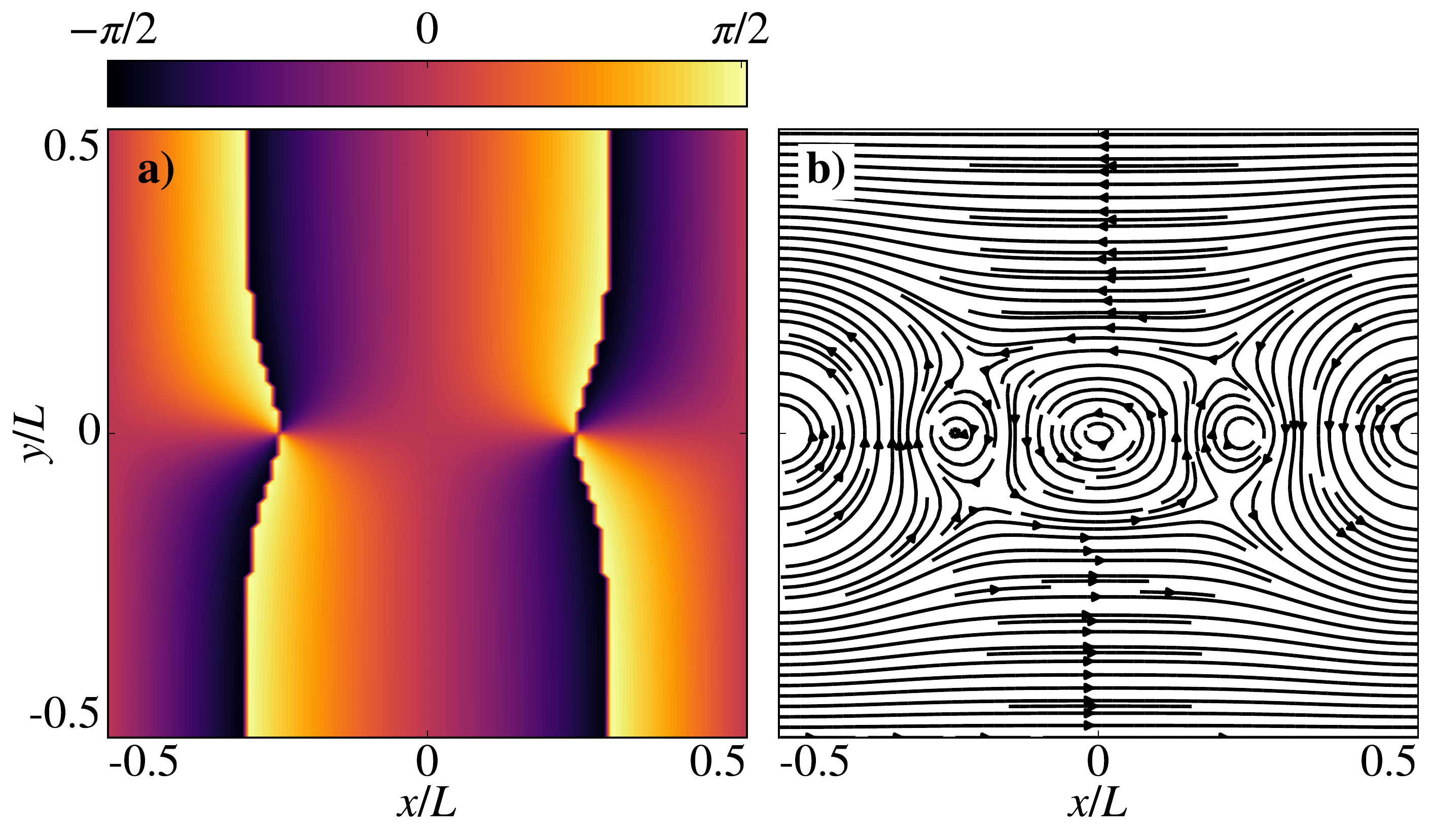}
\caption{Evidence of vortex nucleation. a) The phase of the pair correlation, showing a winding of $2\pi$ around each of the vortices. b) Streamlines of the current density, as given by \cref{eq:current}, which gives its direction at every point, showing that supercurrents circulate around the vortices. }
\label{fig:current}
\end{figure}

The two configurations show markedly different behaviors, as is shown in \cref{fig:vortex} where we set $\alpha=20$ (the results are qualitatively the same for all $\alpha \gg 1$, which corresponds to the magnetization saturating before it reaches the outer edges of the magnetic regions). The uniform exchange field in the parallel configuration introduces a phase shift between the superconductors, so that a net supercurrent flows between them. Otherwise, the system is unaffected. The pair correlation decays towards the center of the TI, but remains nonzero everywhere, as seen in \cref{fig:vortex}a). In the antiparallel configuration, there is no net current due to the antisymmetry of the exchange field, which induces an antisymmetric current density distribution. Furthermore, the exchange field produces a net effective flux $\Phi_h\simeq 2h_0L$, which may cause vortex nucleation.
This is shown in \cref{fig:vortex}b) for $h_0 = 2v_F/\xi$. In this case, two vortices appear along the $x$ axis---the region of largest effective flux density. \cref{fig:vortex}c)-d) shows the spatial distribution of the density of states at zero energy for the two configurations. In the parallel configuration, $N(\vec{r},0)$ is clearly uniformly suppressed throughout the entire system, as is expected due to the presence of a proximity induced energy gap. In the antiparallel configuration on the other hand, the presence of the vortices, which have normal cores, leads to a more complicated topography of the density of states, wherein a normal state value of $N = 1$ is found in localized regions surrounding the vortices. The topological nature of these vortices is illustrated by the phase of the pair correlation, which is shown in \cref{fig:current}a). It is seen that for any closed contour around a vortex, it is necessary to traverse two discontinuous jumps of value $\pi$, giving a total winding of $2\pi$. This is the hallmark of a vortex. Another signature of vortices is circulating supercurrents, as is shown in \cref{fig:current}b), in which streamlines of the current density, as given by \cref{eq:current}, are plotted. Since the eddies produce an out-of-plane magnetic field, which should be detectable using, for instance, a scanning nanoSQUID device~\cite{vasyukov_nn_2013}, this provides means for experimentally verifying the presence of vortices.

The behavior of the vortices is greatly influenced by the symmetries of the system. The model considered herein is symmetric about the $y$ axis, and either symmetric or antisymmetric about the $x$ axis, depending on the applied exchange field. This means that a single vortex pair can only be located on symmetrically opposite sides of the origin, along either the $x$ or the $y$ axis without breaking the symmetries of the system. For an increasing exchange field amplitude, $h_0$, the antiparallel configuration will lead to the appearance of an increasing number of vortices. The vortices enter the system from the vacuum edges, and must do so in pairs from opposite sides. Due to the low flux density near the vacuum edges, even the slightest additional increase in $h_0$ will cause the vortices to translate along the $y$ axis, meet at the origin, and stabilize at a location along the $x$ axis, as shown in \cref{fig:vortex}. As $h_0$ is increased further, vortices accumulate along the $x$ axis. This will, in turn, result in a complete suppression of the density of states in their vicinity, whereas superconductivity will still be present closer to the vacuum edges. Another interesting feature of the inhomogeneous effective flux density is that it leads to significant vortex pinning. Indeed, if the superconducting leads are given a phase difference, for instance by applying a current bias, so that a net supercurrent flows between them, the vortex positions are only slightly perturbed. This is in contrast to the behavior of conventional SNS Josephson weak links with an applied magnetic flux, where a phase difference leads to a transversal shift of the vortex positions \cite{cuevas_prl_2007, bergeret_jltp_2008}. 

\begin{figure}
\centering
\includegraphics[width=\columnwidth]{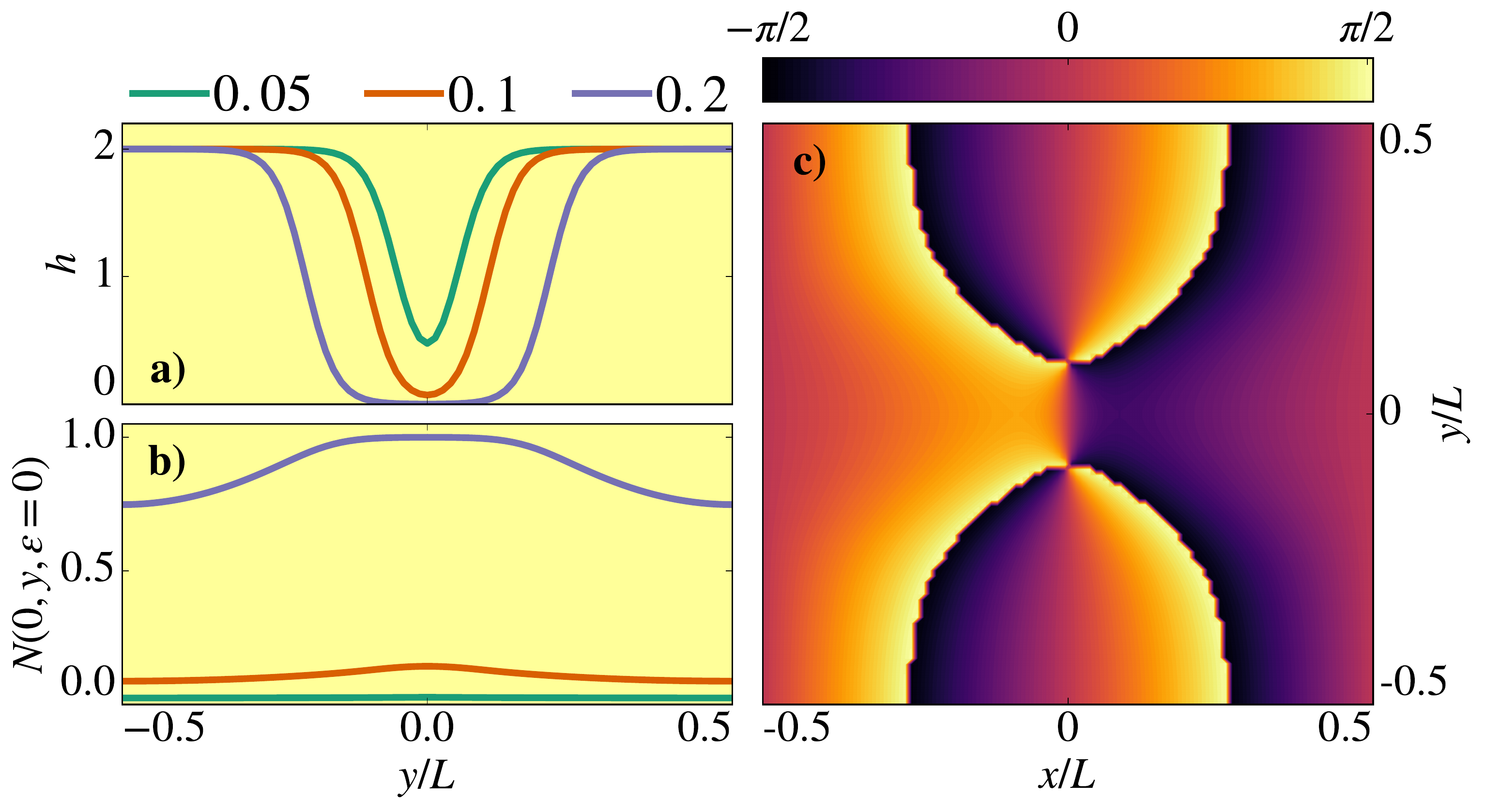}
\caption{An analysis of the effect of the spacer layer. a) The transversal distribution of the exchange field $\vec{h}$ for increasing size $\delta$ of the central region of suppressed magnetization. b) The zero-energy density of states along the $y$ axis for the exchange fields in a). c) The phase of the pair correlation for the case where $\delta = 0.2$, showing the appearance of a vortex--antivortex pair.}
\label{fig:width}
\end{figure}

The exchange field induced on the topological insulator is assumed generated by two separate ferromagnets with an intermediate spacer layer. In the parallel configuration, this will likely create a suppression of the induced exchange field beneath the spacer. 
The resulting $\partial_y h_x \neq 0$ could in itself induce vortices in the system, in addition to the switching-effect we have described above. 
To investigate this, 
%effect, 
we consider a parallel exchange field ${\vec{h}(y) = h_0\left\{1 + 0.5\left(\tanh[\alpha (y/L - \delta)] - \tanh[\alpha (y/L + \delta)]\right)\right\}\hat{x}}$, where $\delta$ is another shape factor indicating the width of the central dip in $\vec{h}(y)$. The exchange field is plotted along the transversal direction $y$ for increasing $\delta$ in \cref{fig:width}a). Since the exchange field is symmetric, the effective flux $\Phi_h = 0$. 
Nonetheless, topological excitations in the form of vortex--antivortex pairs may be induced. It is clear that this can happen if an effective flux greater than $\Phi_0$ passes through any subdomain of the system within which vortex nucleation is allowed by symmetry. The central dip in the exchange field will cause vortices to nucleate where $\nabla\times\vec{h}$ is largest and positive, at $y = \delta L$, whereas antivortices will nucleate at $y = -\delta L$, where the largest negative effective flux density is found. 
To conserve the symmetry of the system, a single vortex--antivortex pair must appear along the $y$ axis. The first appearance of such a pair may therefore be gauged from the zero-energy density of states along this line, as is shown in \cref{fig:width}b). It is seen that $N$ remains gapped for a sufficiently small dip, as exemplified by $\delta = 0.05$ and $\delta = 0.1$. This shows that the vortex spin valve effect is robust against small deviations from a constant exchange field due to the presence of the spacer layer. For $\delta = 0.2$, however, a vortex--antivortex pair appears, and the gap in the density of states closes. This is verified from the phase of the pair correlation, shown in \cref{fig:width}c), where the two vortices along the $y$ axis are seen to have opposite windings.

\emph{Conclusion.} We have considered a Josephson weak link made on the surface of a topological insulator, onto which is proximity coupled two ferromagnets separated by a spacer. By using microscopic calculations, we have shown that it is possible to switch vortices on and off in this system solely by toggling between an antiparallel and parallel configuration of the ferromagnets, respectively. We further show that this vortex spin-valve effect is robust against small deviations in the induced exchange field caused by the spacer layer.      

\begin{acknowledgments}
  \emph{Acknowledgments}.
J. L. and A. S.  acknowledge funding from the Research Council of Norway Center of Excellence Grant Number 262633, Center for Quantum Spintronics. J. L. and M. A. also acknowledge funding from the NV-faculty at the Norwegian University of Science and Technology. A. S. and H. G. H. acknowledge funding from  the Research Council of Norway  Grant Number 250985.  J. L. acknowledges funding from Research Council of Norway Grant No. 240806.  
\end{acknowledgments}

\end{document}